\documentclass[prb,twocolumn,floats,aps]{revtex4}

\usepackage{graphicx}
\begin{document}


\title{Two-orbital Systems with Crystal Field Splitting and Interorbital Hopping}
\author{Yun Song}
\affiliation{Department of Physics, Beijing Normal University,
Beijing 100875, China}
\author{Liang-Jian Zou}
\affiliation{Key Laboratory of Materials Physics, Institute of
Solid State Physics, Chinese Academy of Sciences, P. O. Box 1129,
Hefei 230031, China }
\date{\today}

\begin{abstract}
    The nondegenerate two-orbital Hubbard model is studied
    within the dynamic mean-field theory to reveal the influence of
    two important factors, i.e. crystal field splitting and interorbital
    hopping, on orbital selective Mott transition (OSMT) and realistic
    compound Ca$_{2-x}$Sr$_{x}$RuO$_{4}$. A distinctive feature of the
    optical conductivity of the two nondegenerate bands is found in
    OSMT phase, where the metallic character of the wide band is
    indicated by a nonzero Drude peak, while the insulating narrow
    band has its Drude peak drop to zero in the mean time. We also
    find that the OSMT regime expands profoundly with the increase of
    interorbital hopping integrals. On the contrary, it is shown that
    large and negative level splitting of the two orbitals diminishes the
    OSMT regime completely. Applying the present findings to compound
    Ca$_{2-x}$Sr$_{x}$RuO$_{4}$, we demonstrate that in the doping
    region from $x=0.2$ to $2.0$, the negative level splitting is
    unfavorable to the OSMT phase.


\end{abstract}
\maketitle

\section{Introduction}
\label{intro}

 The orbital degree of freedom is one of the key points to
understand the rich and complicated properties of transition metal
oxides \cite{Nagaosa}, and its roles in Mott-Hubbard metal-insulator
transition (MIT) have attracted much interest in recent years. The
orbital selective Mott transition (OSMT) in multi-orbital correlated
systems, where the Mott-Hubbard metal-insulator transition (MIT) in
different orbitals happens successively with the variation of
doping, interaction, or field, was proposed \cite{Anisimov} to
interpret the two consecutive phase transitions and spin state
transition in the layered perovskite Ca$_{2-x}$Sr$_{x}$RuO$_{4}$
\cite{Nakatsuji}. Besides, it was also applied to other orbital
compounds, such as V$_{2}$O$_{3}$ and VO$_{2}$ \cite{Laad}. However,
this OSMT scenario has not been verified by experiments in the OSMT
candidate Ca$_{2-x}$Sr$_x$RuO$_4$ till now. The optical spectra
\cite{Lee} and the angle-resolved photoelectron spectroscopy (ARPES)
\cite{Wang} experiments were performed to examine the electronic
states and spectral properties of Ca$_{2-x}$Sr$_x$RuO$_4$, but
neither considerable contraction of Fermi surface volume \cite{Lee}
nor the OSMT gap for narrow $d_{xz/yz}$ bands \cite{Wang} was found
near $x=0.5$. Whereas the most recent result obtained by ARPES
experiment on Ca$_{2-x}$Sr$_x$RuO$_4$ at $x=0.2$ presented a
different scenario, where the low-energy quasiparticle excitations
in wide $d_{xy}$ orbital disappeared completely \cite{Neupane}.

Great theoretical efforts have been paid to unearth the essential of
the OSMT by the dynamical mean-field theory (DMFT) and some other
approaches
\cite{Koga,Liebsch,Knecht,Medici,Arita,Ferrero,Ruegg,Dai}. It has
been confirmed that the two successive MITs and the OSMT are robust
features in the two-orbital and three-orbital Hubbard models with
asymmetric bandwidths. Meanwhile, the essential aspects of the OSMT
are not well understood. For example, the effect of Hund's coupling
in OSMT is still not clear. Koga {\it et al.} \cite{Koga} and
Liebsch {\it et al.} \cite{Liebsch} found that the full Hund's
coupling is crucial for the existence of OSMT in two-orbital Hubbard
model with asymmetric bandwidth using the DMFT with quantum Monte
Carlo solver and exact diagonalization solver, respectively.
However, utilizing the slave spin mean-field theory and DMFT, Medici
{\it et al.} \cite{Medici} suggested that the OSMT exists over a
wider range of $J$ when the ratio of the narrow and wide bandwidths
is small enough, even in the full SU(4) spin-orbital symmetric case
with $J=0$. On the other hand,  Liebsch {\it et al.} \cite{Liebsch2}
recently introduced a three-orbital Hubbard model with positive
crystal field splitting for the real compound
Ca$_{2-x}$Sr$_x$RuO$_4$, and found a metallic-band insulator
transition for the completely filled $d_{xy}$ orbital and a Mott
transition in the remaining half-filled $d_{xz/yz}$ bands,
precluding the possibility of the OSMT phase in
Ca$_{2-x}$Sr$_x$RuO$_4$.

In order to get a clear picture about the unconventional MITs in
compound Ca$_{2-x}$Sr$_{x}$RuO$_{4}$, all major essentials of the
real compound should be taken into consideration when a realistic
model is set up. In layered perovskite Ca$_{2-x}$Sr$_{x}$RuO$_{4}$,
the nearest-neighbor ($NN$) interorbital hopping integrals are
comparable with that of the intraorbital integrals in the order of
magnitude. However, in previous studies
\cite{Koga,Liebsch,Knecht,Medici,Arita,Ferrero,Ruegg,Dai,Liebsch2},
the $NN$ interorbital hopping integrals were neglected so as to make
the self-energy and Green's function be diagonal in orbital space.
On the other hand, only the hybridization of the two orbitals at the
same site was considered recently in
reference [11,17]. It is expected that the presence of the
interorbital hopping will profoundly affect OSMT. Moreover, both the
experiments \cite{Lee,Wang} and the first principle calculations
\cite{Fang0} demonstrated that with the increase of doping in
Ca$_{2-x}$Sr$_{x}$RuO$_{4}$, the structure distortions arising from
the rotation and tilting of RuO$_{6}$ octahedra play an important
role in the two successive MITs. As we will analyze later, such
distortions lead to the decline of the level center of the narrow
band with respect to the wide one, giving rise to a negative crystal
field splitting.

The aim of this paper is to elucidate the combining effect of two
important factors, i.e. interorbital hopping and crystal field
splitting, on the optical conductivity and the OSMT in multiorbital
correlated model and realistic multiorbital material
Ca$_{2-x}$Sr$_{x}$RuO$_{4}$. We study the optical conduction of
two-orbital Hubbard model within the framework of the extended
linearized dynamical mean field theory (ELDMFT) \cite{Potthoff},
which has been proved extensively to be an efficient method,
especially for the complicated multiorbital models. We demonstrate
how the interorbital hopping and the negative crystal field
splitting affect the nature of OSMT and the optical conductivity of
the systems. We find that the OSMT regime enlarges in the presence
of the interorbital hopping, but the negative crystal field
splitting strongly restrains the occurrence of OSMT. In addition,
the unconventional MITs in compound Ca$_{2-x}$Sr$_{x}$RuO$_{4}$ is
also discussed. The rest of the paper is organized as follows. In
Sec.~\ref{mod}, we describe the model Hamiltonian and outline the
ELDMFT method; in Sec.~\ref{res}, we present the major theoretical
results of the interorbital hopping and the crystal field splitting
on the OSMT, and the possible application on compound
Ca$_{2-x}$Sr$_{x}$RuO$_{4}$. Sec.~\ref{con} is devoted to the
conclusions.

\section{Model and Methods}
\label{mod}

  We start from the two-orbital Hubbard model,
\begin{eqnarray}
H&=&-\sum_{\langle ij\rangle} \sum_{ll^{\prime}\sigma}
t_{ll^{\prime}} C^{+}_{il\sigma}C_{jl^{\prime}\sigma}
     +\sum_{il\sigma}(\epsilon_{l}-\mu)C^{+}_{il\sigma}C_{il\sigma}
\nonumber \\
         &&+ \frac{U}{2}\sum_{il\sigma}n_{il\sigma}n_{il\bar{\sigma}}
         +\frac{J}{2}\sum_{i,l\neq l^{\prime},
           \sigma}C^{+}_{il\sigma}C^{+}_{il\bar{\sigma}}
          C_{il^{\prime}\bar{\sigma}}C_{il^{\prime}\sigma}
\nonumber\\
        &&+\frac{U^{\prime}}{2}\sum_{i,l\neq
        l^{\prime}}n_{il}n_{il^{\prime}}
         +\frac{J}{2}\sum_{i,l\neq l^{\prime},\sigma\sigma^{\prime}}
         C^{+}_{il\sigma} C^{+}_{il^{\prime}\sigma^{\prime}}
         C_{il\sigma^{\prime}}C_{il^{\prime}\sigma}.
\nonumber\\
\label{Eq.1}
\end{eqnarray}
Here index $l$ ($l^{\prime}$) = $a$ or $b$ refers to the narrow or
wide orbital, and $t_{aa}$ ($t_{bb}$) and $t_{ab}$ denote the $NN$
intraorbital and interorbital hopping integrals, respectively; $U$
and $U^{\prime}$ represent the on-site intraorbital and the
interorbital Coulomb repulsions between electrons; and $J$ denotes
the Hund's coupling. The crystal field splitting arising from the
structure distortions is defined as
$\epsilon_{d}=\epsilon_{a}-\epsilon_{b}$.

In the presence of the finite $t_{ab}$, we introduce a canonical
transformation \cite{Song},
\begin{eqnarray}
C_{ia\sigma}&=&u\alpha_{i\sigma}+v\beta_{i\sigma}
\nonumber\\
C_{ib\sigma}&=&-v\alpha_{i\sigma}+u\beta_{i\sigma}. \nonumber
\end{eqnarray}
Where $u$ and $v$ are the coefficients of the canonical
transformation; $\alpha$ and $\beta$ represent fermion annihilation
operators of the new effective orbitals, with only intraorbital
hoppings $t_{\alpha}=t_{aa}u^2+t_{bb}v^2-t_{ab}uv$ and
$t_{\beta}=t_{aa}u^2+t_{bb}v^2+t_{ab}uv$, and vanishing interorbital
hoppings. As a result, the two-orbital model with interorbital
hopping (Eq.~(\ref{Eq.1})) is transformed into a new effective
two-orbital model without interorbital hopping,
\begin{eqnarray}
H&=&-t_{\alpha}\sum_{\langle
ij\rangle,\sigma}\alpha^{+}_{i\sigma}\alpha_{j\sigma}
-t_{\beta}\sum_{\langle
ij\rangle,\sigma}\beta^{+}_{i\sigma}\beta_{j\sigma}
\nonumber\\
&&+(\epsilon_{a}-\mu)\sum_{i\sigma}(\alpha^{+}_{i\sigma}\alpha_{i\sigma}
+\beta^{+}_{i\sigma}\beta_{i\sigma}) \nonumber\\
&&-\epsilon_{d}\sum_{i\sigma}(v^2\alpha^{+}_{i\sigma}\alpha_{i\sigma}
-uv(\alpha^{+}_{i\sigma}\beta_{i\sigma}
+\beta^{+}_{i\sigma}\alpha_{i\sigma})
+u^2\beta^{+}_{i\sigma}\beta_{i\sigma})
\nonumber\\
&&+H^{eff}_{int}. \label{Eq.3}
\end{eqnarray}
Here $H^{eff}_{int}$ is the interaction Hamiltonian under the
canonical transformation. From Eq.(2), we can distinguish the
effects of the interorbital hopping from the crystal-field splitting
more clearly within the new fermion representation. That is, the
finite interorbital hopping $t_{ab}$ increases the bandwidth ratio
of the two effective orbitals; while the crystal field splitting
$\epsilon_d$ induces a certain extent of the orbital hybridization
of the two effective orbitals at the same site. When
$\epsilon_{d}$=0, the Green's function is diagonal in the new
fermion representations, and we can employ the DMFT approach to
study the two-orbital model easily. On the other hand, when both the
crystal field splitting and the interorbital hopping are taken into
account, the off-diagonal components of the Green's function appear
in this fermion representation. In this situation, to discuss the
effect of crystal field splitting, we can use the similar technique
adopted by Medici {\it et al.} \cite{Medici} and Buenemann {\it et
al.} \cite{Buenemann} for the two-orbital models with the on-site
orbital hybridization. In addition, further transformation can be
introduced to diagonal the Green's function. More details of our
method are given in Sec.~\ref{sub3}.

In the DMFT approach, the self-consistent loop is built to fulfil
two equations,
\begin{equation}
G^{l}_{imp}(\omega)=G^{l}(\omega) \label{eq:sc1}\nonumber
\end{equation}
and
\begin{equation}
\Sigma^{l}_{imp}(\omega)=\Sigma^{l}(\omega).\label{eq:sc2}\nonumber
\end{equation}
Where $G^{l}_{imp}(\omega)$ and $\Sigma^{l}_{imp}(\omega)$ represent
respectively the Green's function and the self-energy of the
Anderson impurity model mapped from the two-orbital Hubbard model.
We adopt the ELDMFT approach \cite{Potthoff}, which is a fast and
efficient method for the multi-orbital systems. The result of ELDMFT
for one-band Hubbard model has been compared with that obtained by
the exact diagonalization \cite{Georges} and numerical
renormalization group \cite{Bulla} solvers, and a good agreement has
been achieved in a rather wide $U$ range.

In the DMFT approach with the exact diagonalization and numerical
renormalization group solvers, it has been proved that the lower and
higher energy behaviors of the self-energy play the crucial roles.
In order to facilitate the calculation and to consider the lower and
higher frequency features properly, the ELDMFT approach introduces
two self-consistent equations \cite{Potthoff},
\begin{equation}
V_l^2=z_l\sum_{j\neq i}t_{ij}^2 \label{eq:ELsc1} \nonumber
\end{equation}
and
\begin{equation}
n^l=n^l_{imp}.\label{ELeq:ELsc2}\nonumber
\end{equation}
Where $V_l$ is the hybridizing parameter in the Anderson model
\cite{Potthoff,Georges,Bulla}, $z_l=1/(1-d\Sigma^l(0)/d\omega)$ is
the quaiparticle weight, and $n^l$ ($n^l_{imp}$) represents the
occupation number of the lattice (impurity) model. In addition, the
free-particle DOS is a semicircular type. These two equations
correspond to those of the conventional DMFT approach in
Eq.~(\ref{eq:sc2}), respectively, and there is no other extra
restriction employed in ELDMFT. The ELDMFT method is able to access
the entire parameter region since the Luttinger theorem $\langle
N\rangle =nL$ ($L$ is the number of lattice size) is exactly
fulfilled in the whole range of filling.

In the new fermion representation obtained by the canonical
transformation \cite{Song}, the optical conductivity of each
effective orbital in the paramagnetic system without the
interorbital hopping reads \cite{Georges,Rozenberg}
\begin{eqnarray}
 \sigma_{l}(\omega)&=&\pi\int d\omega' d\epsilon
       D^{l}(\epsilon)\rho_{\epsilon}^{l}(\omega')
        \rho_{\epsilon}^{l}(\omega'+\omega)
        \nonumber\\
      &\times&\frac{f(\omega')-f(\omega'+\omega)}{\omega},
\end{eqnarray}
where $D^{l}(\epsilon)$ and $\rho^{l}(\omega)$
($\rho^l(\omega)=-\frac{1}{\pi}Im G^l(\omega+i\gamma)$) denote the
transition matrix elements and the density of states (DOS) in the
new orbital channel $\alpha$ or $\beta$, respectively; and
$f(\omega)$ is the Fermi-Dirac distribution function. A small
spectra broadening ($\gamma$=0.1) is introduced for the calculations
of DOS. When there is no hopping between the new quasiparticle
channels, the total optical conductivity $\sigma(\omega)$ is the
summation of the contributions of two quasiparticle orbitals,
$\sigma_{\alpha}(\omega)$ and $\sigma_{\beta}(\omega)$.
In the presence of interorbital hopping, the full optical
conductivity for the Green's functions and the transition matrix
$D^{ll'}(\epsilon)$ with off-diagonal terms can be obtained by,
\begin{eqnarray}
        \sigma(\omega)&=&\pi\int d\omega' d\epsilon
       \sum_{l,l'} D^{ll'}(\epsilon)\rho_{\epsilon}^{l}(\omega')
        \rho_{\epsilon}^{l'}(\omega'+\omega)
        \nonumber\\
      &\times&\frac{f(\omega')-f(\omega'+\omega)}{\omega}.
\end{eqnarray}
In this paper, we mainly focus on the optical conductivity of each
orbital, $\sigma_{l}$, so as to elucidate the metallic or insulating
nature of each orbital.

In many transition-metal oxides, the transition-metal ions and
oxygen ions usually form TMO$_{6}$ octahedra. The TMO$_{6}$ clusters
distorted from the perfect octahedra contribute the crystal field
splitting. Starting from the single-particle potential $V(\vec{r})$,
one can estimate the splittings in the point-charge approximation of
the crystalline field theory \cite{Sugano}. Expanding the
single-particle potential $V(\vec{r})$ $=$
$\sum_{j}q_{j}/|\vec{R_{j}}-\vec{r}|$ in terms of the spherical
harmonics:
\begin{equation}
  V(\vec{r})=\sum^{\infty}_{k=0}\sum^{k}_{m=-k}A_{km}r^{k}Y_{km}(\theta,\phi)
\end{equation}
where $A_{km}
=\frac{4\pi}{2k+1}\sum_{j}\frac{q_{j}}{R_{j}^{k+1}}Y_{km}^{*}
(\theta_{j},\phi_{j})$. The contribution of the crystal field
Hamiltonian, H$_{CF}$, reads
\begin{equation}
  H_{CF} =\sum_{\sigma}\int d\vec{r} \psi_{\sigma}^{+}(\vec{r})V(\vec{r})
           \psi_{\sigma}(\vec{r})d\vec{r}.
\end{equation}
In the second quantization representation, H$_{CF}$ can be rewritten
as
\begin{equation}
  H_{CF} =\sum_{j}\sum_{ll'\sigma}V_{ll'}C^\dag_{jl\sigma}C_{jl'\sigma}
\end{equation}
where
\begin{equation}
V_{ll'}=\int w_{j}^{l}(\vec{r})V(\vec{r})
w_{j}^{l'}(\vec{r})d\vec{r},\nonumber
\end{equation}
with
$\psi_{\sigma}(\vec{r})=\sum_{l}\sum_{j}w_{j}^{l}(\vec{r})c_{jl\sigma}$
and $w_{j}^{l}$ being the Wannier basis consists of the
wavefunctions of the transition-metal $d$ orbits and of the
nearest-neighbor atomic orbits. In the high symmetric situation, the
matrix $V_{ll'}$ is diagonal, giving rise to $\epsilon_{l}=V_{ll'}$.
Later, we will evaluate the crystal field splitting based on these
equations and the experimental data of the crystal structures of the
family compounds Ca$_{2-x}$Sr$_{x}$RuO$_{4}$.

\section{Results}
\label{res}

    We study the effects of interorbital hopping and crystal field
splitting separately in the asymmetric model, and discuss the
application to real compound Ca$_{2-x}$Sr$_{x}$RuO$_4$. The
calculations are performed for the Bethe lattice with semicircular
DOS, and our results are measured in units of the hopping integral
$t_{bb}$. In our calculation, the bandwidthes of the free-particle
spectra are fixed as $W_a=2$ and $W_b=4$.

\begin{figure}
\includegraphics[viewport= 0 0 500 300, width=\columnwidth, clip ]{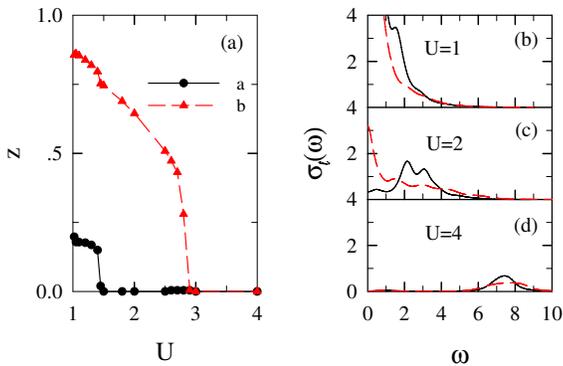}
\caption{(Color online) (a) Quasiparticle weight as a function of
interaction $U$; (b)-(d) evolutions of optical conductivity in
different orbital channel with U from metallic (b), OSMT (c), to
insulating (d) phases in the two-orbital half-filling Hubbard model
with $J=0.5$, $t_{aa}=0.5$, $t_{bb}=1.0$, $t_{ab}=0$, and
$\epsilon_{d}$=0.} \label{fig_1}
\end{figure}

\subsection{OSMT and optical conductivity}
\label{sub1}

    To present the distinction of the optical conductivity in
different phases, we first investigate the simplest two-orbital
model without interorbital hopping ($t_{ab}=0$) and crystal field
splitting ($\epsilon_d=0$). The quasiparticle weight $Z$ is a
powerful tool to characterize the MIT and the OSMT. We present the
evolution of quasiparticle weight with the interaction $U$ in
Fig.~\ref{fig_1}(a) for a two-orbital model with different orbital
bandwidths $W_1/W_2=0.5$. We find that when the Coulomb interaction
$U$ increases to $U_{c_1}=1.45$, the first MIT, i.e. the OSMT,
happens. The second transition occurring at $U_{c_2}=2.90$ is the
conventional MIT.

The quasiparticle contribution to the optical conductivity arises
from the electronic transitions between the lower and the upper
Hubbard subbands. As shown in Fig.~\ref{fig_1}(b), the Drude peaks
of both orbitals are considerably large as $U<U_{c_1}$, which
demonstrates that the system is in the metallic phase. On the
contrary, the Drude peaks of both ortitals vanish completely in the
insulating region with $U> U_{c_2}$, as shown in
Fig.~\ref{fig_1}(d). However, in the OMST regime with $U_{c_1}<U<
U_{c_2}$, the Drude peak of the narrow band vanishes and only the
wide band contributes to the conduction channel in the optical
conductivity, as shown in Fig.~\ref{fig_1}(c). Similar to the
orbital ordering in manganites and vanadates, the OSMT phase is also
orbital polarized in the wide channel \cite{Song2}. Therefore, the
OSMT phase can be detected by using the polarized synchrotron
radiation light to measure the distinct optical conduction behaviors
of different orbitals.

Apart from the low-energy Drude-like feature coming from the
coherent quasiparticle DOS existing near $E_F$, there is another
spectral feature, as shown in Fig. 1(b)-(d), centered at the
intermediate energy, which represents the charge transfer from the
lower and upper Hubbard subbands. Since the general feature of DOS,
i.e. the lower and upper Hubbard subbands and a quasiparticle
resonance at Fermi surface, can be reproduced by ELDMFT, our
numerical results of optical conductivity are qualitatively
accurate. To quantitatively compare with experiments on
Ca$_{2-x}$Sr$_x$RuO$_4$, more accurate DOS should be adopted rather
than the semicircular DOS, and all three $t_{2g}$ orbitals should be
taken into account. This needs to be studied in future work.

\begin{figure}
\includegraphics[viewport= 0 0 500 300, width=\columnwidth, clip ]{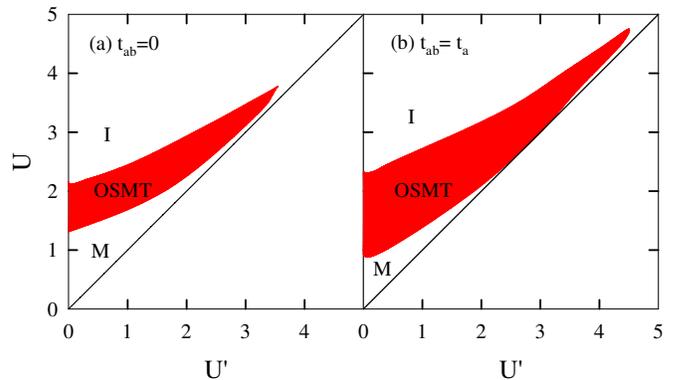}
\caption{(Color online) Phase diagram of the two-orbital Hubbard
model (a) without interorbital hopping, and (b) with interorbital
hopping. Theoretical parameters: $J=0.5$, $t_{aa}=0.5$,
$t_{bb}=1.0$, and $\epsilon_{d}$=0.} \label{fig_2} \label{fig_2}
\end{figure}

\subsection{Effect of interorbital hopping}
\label{sub2}

Since the interorbital hopping is significant in real material, it
is crucial to dig out its effect on the phase diagram and the
optical conductivity. To our knowledge, no work has been done to
take this important factor into consideration. Till now, only the
hybridization of the two orbitals at the same site had been studied
by DMFT \cite{Buenemann}, where the OSMT tends to be destroyed.
However, the physical origins of the on-site intra-atomic
hybridization and NN interorbital hopping are quite different. By
analyzing the quasiparticle weight at Fermi surface, we obtain the
phase diagram of the two-orbital Hubbard model with interorbital
hopping integrals $t_{ab}=0.5$, as shown in Fig.~\ref{fig_2}(b). For
the convenience of comparison, the phase diagram for $t_{ab}=0.0$ is
also shown in Fig.~\ref{fig_2}(a). In both cases, a stable OSMT
regime (the red shadow) lies between the Mott insulating (denoted by
"I") and the conventional metallic (denoted by "M") phases. When the
interorbital hopping is finite, for example $t_{ab}=0.5$, the area
of the OSMT regime greatly expands, in comparison with the case
$t_{ab}=0$.

  It should be noted that the OSMT we observed here is for the
effective orbitals in the new fermion representation, which are the
hybridized orbitals of the two realistic orbitals \cite{Song}.
Within this representation, one could find that the bandwidth ratio
of the current two effective orbitals increases with the increase of
the interorbital hopping integrals $t_{ab}$: the wide quasiparticle
band becomes wider, while the narrow one becomes narrower. Thus the
OSMT occurs more easily, and its area expands a lot in the phase
diagram for large $t_{ab}$. In addition, we also find that for zero
and nonzero interorbital hopping, the OSMT regime is close to the
$U=U'$ line, but never crosses it. Our result of t$_{ab}$=0 agrees
with Medici {\it et al.}'s phase diagram obtained by slave boson
approach; however, at $J=0.5$, our result is slightly different from
the result obtained by Koga {\it et al.} \cite{Koga} for $J=0$ in
which the OSMT regime crosses the $U=U'$ line.

\begin{figure}
\includegraphics[viewport= 0 0 450 280, width=\columnwidth, clip ]{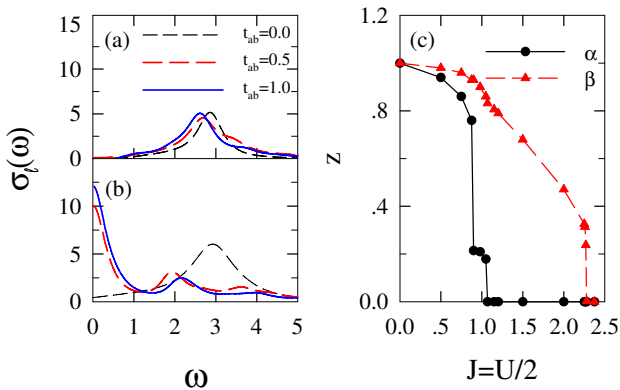}
\caption{(Color online) The optical conductivities of the narrow
orbital (a) and of the wide orbital (b) as functions of energy for
different interorbital hopping integrals: $t_{ab}$=0, 0.5, and 1.0
as $U=2.25$; (c) Quasiparticle weight as a function of $J$ with
$t_{ab}=t_{a}$. Theoretical parameters: $U'=0$, $J=U/2$,
$t_{aa}=0.5$, $t_{bb}=1.0$, and $\epsilon_{d}$=0.} \label{fig_3}
\end{figure}

   The important role of the interorbital hopping $t_{ab}$ in the
OSMT also manifests in the optical conductivity. Starting from an
insulating case with $U=2.25$, $U'=0$ and $t_{ab}=0$, we obtain the
typical feature of insulating optical conductivity for both the
narrow band and the wide one. Without the interorbital hopping, the
system is a conventional Mott insulator, and the Drude peaks of both
effective orbitals vanish. As shown by the black dashed lines in
Fig.~\ref{fig_3}(a) and \ref{fig_3}(b), there is no Drude peak for
both $\sigma_{\alpha}(\omega)$ and $\sigma_{\beta}(\omega)$, and the
peaks in the intermediate energy region clearly show the hopping
between the subbands separated by the intraorbital Coulomb
interactions. It is obvious that the interorbital hopping $t_{ab}$
does not significantly change the intermediate-energy band
structures in the DOS of the narrow band. However, in
Fig.~\ref{fig_3}(b), we find that the Drude peak of the optical
conductivity of the wide band $\sigma_{\beta}(\omega)$ crucially
changes with the increase of the interorbital hopping $t_{ab}$. As
$t_{ab}$ increases to 0.5 and unity, the Drude peak in
$\sigma_{\beta}(\omega)$ considerably increases, implying that the
system enters into the OSMT regime. We also find that the
quasiparticle weight drops abruptly at the two successive MIT
points, as shown in Fig.~\ref{fig_3}(c), when the finite
interorbital hopping is taken into account. We observe a small
plateau near the metal-OSMT transition point, which primarily shows
that both of the two successive MITs are first-order. The order of
the OSMT deserves further study.

\subsection{Effect of crystal field splitting}
\label{sub3}

The level splitting of the two orbitals $\epsilon_d$ arising from
the crystal field is another important factor that influences the
essential of the MIT and the OSMT in the two-orbital systems. To
study the evolution of the optical conductivity of the compound
Ca$_{2-x}$Sr$_{x}$RuO$_{4}$ with the crystal field splitting
$\epsilon_d$, we start from an OSMT phase with the parameters
$U=2.0$, $U'=1.0$ and $J=0.5$, which lies in the red shadow regime
in the phase diagram shown in Fig.~\ref{fig_2}(b).

As shown in Eq.~(\ref{Eq.3}), there are hybridization terms
$\alpha^{+}_{i\sigma}\beta_{i\sigma}$ and
$\beta^{+}_{i\sigma}\alpha_{i\sigma}$ for the two effective orbitals
on the same lattice when both the interorbital hopping and crystal
field splitting are nonzero. In this condition, the Green's function
is defined as,
\begin{equation}
{\bf G}_0\left(\omega,k\right)^{-1}=\left(
\begin{array}{cc}
\omega+\mu-\epsilon_{\alpha}( k) & \epsilon_duv\\
\epsilon_duv & \omega+\mu-\epsilon_{\beta}( k)
\end{array}
\right), \label{Eq_G0}
\end{equation}
where $\epsilon_\alpha (k)$ and $\epsilon_\beta (k)$ are the bare
dispersion relations for the $\alpha$ and $\beta$ effective
orbitals, respectively. Obviously, we can directly define the local
Green function as
\begin{equation}
G^{\alpha (\beta)}(\omega)=\int^{\infty}_{-\infty} dx
\frac{\rho_0(x)}{\omega+\mu-x-\Sigma^{\alpha
(\beta)}\left(\omega\right)-\Delta^{\alpha (\beta)}
\left(\omega,x\right)},
\end{equation}
with
\begin{equation}
  \Delta^{\alpha (\beta)}\left(\omega,
     x\right)=\frac{\left(\epsilon_duv
     +\Sigma^{\alpha\beta}\right)^2}
     {\omega+\mu-x-\Sigma^{\beta (\alpha)}(\omega)}.
\nonumber\\
\end{equation}
Here $\Sigma^{\alpha (\beta)}(\omega)$ and $\Sigma^{\alpha
\beta}(\omega)$ represent the diagonal and off-diagonal terms of the
self-energy matrix, respectively.

Based on the special feature of the above off-diagonal Green's
function in Eq.~(\ref{Eq_G0}), we can introduce an unitary matrix
$U(\omega,k)$ to diagonalize Eq.~(\ref{Eq_G0}) as an effective
Green's function,
\begin{eqnarray}
\widetilde{{\bf G}}_0\left(\omega,k\right)^{-1}&=&U{\bf
G}_0\left(\omega,k\right)^{-1}U^{-1}
\nonumber\\
&=&\left(
\begin{array}{cc}
\omega+\mu-\epsilon^{+}( k) & 0\\
0 & \omega+\mu-\epsilon^{-}( k)
\end{array}
\right), \label{Eq_Gwan}\nonumber\\
\end{eqnarray}
with
\begin{equation}
\epsilon^{\pm}(k)=\frac{1}{2}(\epsilon_{\alpha}(k)
+\epsilon_{\beta}(k)\pm\sqrt{(\epsilon_{\alpha}(k)
-\epsilon_{\beta}(k))^2+4\epsilon_d^2u^2v^2}).
\end{equation}
For a half-filling asymmetric two-orbital system, when the
relationship $|\epsilon_{\alpha}(k)
-\epsilon_{\beta}(k)|<<|\epsilon_d|$ is satisfied in the energy
region close to the Fermi surface, we can approximate the above
dispersion relations in the first order as,
\begin{eqnarray}
\epsilon^{\pm}(k)&=&\frac{1}{2}[\epsilon_{\alpha}(k)
+\epsilon_{\beta}(k)\nonumber\\
&&\pm (2|\epsilon_duv|+\frac{(\epsilon_{\alpha}(k)
-\epsilon_{\beta}(k))^2}{4|\epsilon_duv|})].
\end{eqnarray}
It is obvious that the finite crystal field splitting mixes the two
effective $\alpha$- and $\beta$-orbital in the low energy regime,
and as a result, the OSMT is suppressed.

\begin{figure}
\includegraphics[viewport= 0 0 400 200, width=\columnwidth, clip ]{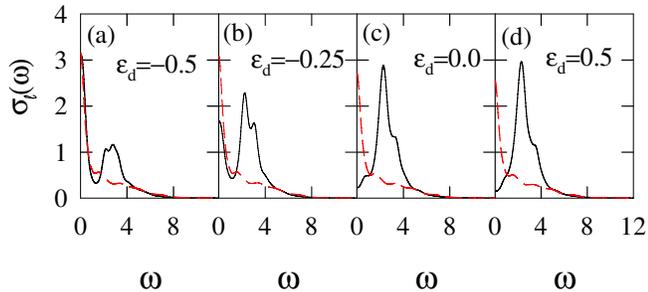}
\caption{(Color online) Effect of crystal field splitting on optical
conductivity of each orbital and OSMT phase in two-orbital Hubbard
model with $U=2.0$, $U'=1.0$, $J=0.5$, $t_{aa}=0.5$, $t_{bb}=1.0$,
and $t_{ab}=0.5$. The black solid line denotes the effective narrow
orbital, and the red dash line the effective wide obital.}
\label{fig_4}
\end{figure}

We employ the ELDMFT approach \cite{Potthoff} to numerically solve
the corresponding impurity problem, and find that the OSMT phase is
strongly influenced by the transfer of electrons between the two
effective orbitals. We start from an OSMT phase with
$\epsilon_d=0.0$, in which the electrons in the narrow orbital are
localized and itinerant in the wide one, respectively. As shown in
Fig.~\ref{fig_4}(c), the OSMT character clearly manifests in the
Drude peak of each orbital: the Drude peak vanishes completely for
insulating narrow orbital, while the wide orbital is metallic with a
finite Drude peak.

When the crystal field splitting becomes negative as
$\epsilon_d=-0.5$ (see Fig.~\ref{fig_4}(a)) and -0.25 (see
Fig.~\ref{fig_4}(b)), the narrow orbital shifts to the lower energy
region, and a fraction of electrons in the wide orbital transfers to
the narrow orbital. The electron filling in the narrow orbital is
more than a half, thus the narrow orbital also becomes metallic. 
In this situation, the Drude peak of the narrow orbital appears
again. On the other hand, as the crystal field splitting is
positive, the narrow orbital keeps its insulating character and the
wide one is metallic. So the system remains in the OSMT phase, as
shown in Fig.~\ref{fig_4}(d).

Furthermore, we find that in both the metallic and insulating phases
near the border of the OSMT regime, a small negative crystal field
splitting can drive neither the metallic system nor the insulating
system into the OSMT phase. Therefore, the negative crystal field
splitting is not in favor of the stable OSMT phase, even in the
presence of the finite interorbital hopping integrals. This result
is contrary to that of Dai {\it et al.} \cite{Dai}, who supported
that the OSMT exists in the three-orbital Hubbard model with crystal
field splitting.

\subsection{Application to Ca$_{2-x}$Sr$_{x}$RuO$_{4}$}
\label{sub4}

\begin{figure}
\includegraphics[viewport= 0 0 500 320, width=\columnwidth, clip ]{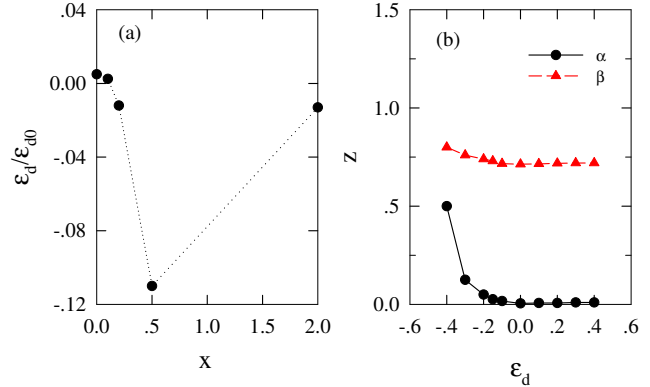}
\caption{(Color online) (a) Crystal field splitting $\epsilon_d$ as
a function of $x$ in compound Ca$_{2-x}$Sr$_{x}$RuO$_{4}$;
(b)Quasiparticle weight as a function of $\epsilon_d$. Other
parameters are the same to Figure 4.} \label{fig_5}
\end{figure}

For the compound Ca$_{2-x}$Sr$_{x}$RuO$_{4}$, the crystal field
splitting arises from the structure distortions, and it can be
approximately modeled as a two-orbital system since the double
degeneracy of the narrow d$_{xz/yz}$ bands is kept over the whole
doping range. Within the framework of the single-ion crystal field
theory \cite{Sugano}, we investigate the crystal field splitting
between the d$_{xz/yz}$ orbitals and the d$_{xy}$ orbital with the
doping concentration x varying from 0.0 to 2.0. According to the
crystal structures for x=2.0, 0.5, 0.2, 0.1 and 0 \cite{Freidt}, we
find that in the most doping region, the narrow d$_{xz/yz}$ bands
are lower than the broad d$_{xy}$ band, showing that the negative
crystal field splitting is dominant. Noting that though the
single-ion crystal field theory is relatively simple, it is
qualitatively correct for the sign and the relative magnitude of the
crystal field splittings. As shown in Fig.~\ref{fig_5}(a), the
averaged splittings between the d$_{xz/yz}$ and the d$_{xy}$ bands
are $\epsilon_{d}/\epsilon_{d_0}$= -0.013, -0.11, -0.012, 0.0025 and
0.005 for x=2.0, 0.5, 0.2, 0.1 and 0, respectively. Here
$\epsilon_{d_0}$ is a positive constant, representing the
$4d$-orbital radial integral of the single-particle potential
($\epsilon_{d_0} = \int_{0}^{\infty} r^2 d{\bf r}
\langle\phi(r)|V(r)|\phi(r)\rangle$). The level separation of the
narrow d$_{xz/yz}$ bands from the wide d$_{xy}$ band in
Ca$_{2-x}$Sr$_{x}$RuO$_{4}$ varies gradually from positive to
negative with the increase of Sr concentration, and have the largest
value as $x=0.5$. This largest level separation arises from the
rotation-type distortion \cite{Freidt}, which becomes the most
significant when $x=0.5$. As $x$ further decreases, the tilting
distortion of the RuO$_{6}$ octahedra \cite{Freidt} becomes more
important, which usually mixes different orbitals.

Since the normal state of Sr$_{2}$RuO$_{4}$ is metallic, our
preceding results shows that negative crystal field splitting is
unfavor of stable OSMT phase. Therefore, our results exclude the
possibility of the OSMT occurring in Ca$_{2-x}$Sr$_{x}$RuO$_{4}$
when x increases from 0.2 to 2.0. Very recently, Liebsch {\it et
al.} \cite{Liebsch2} suggested that in compound
Ca$_{2-x}$Sr$_{x}$RuO$_{4}$, the splitting of the $d_{xz/yz}$ to the
$d_{xy}$ bands is positive. In their study, the band structures in
compound Ca$_{2-x}$Sr$_{x}$RuO$_{4}$ is rigid, and thus the
variation of electron occupations in different orbitals arises from
the change of the positive crystal field splitting. Based on this
hypothesis, Liebsch {\it et al.} \cite{Liebsch2} and Werner {\it et
al.} \cite{Millis} discussed the role of the positive crystal field
splitting on the MIT in two-orbital Hubbard model. However, Lee {\it
et al.} \cite{Lee} have demonstrated that with the doping increasing
from x=0 to 2, the bandwidths of the $d_{xy}$ and $d_{xz/yz}$
orbitals considerably change, implying that the rigid band
approximation is not proper.

Based on the above findings, the two successive MIT in
Ca$_{2-x}$Sr$_{x}$RuO$_{4}$ is obviously independent of the OSMT. A
question arises: what happens in the first unconventional MIT at
x=0.2? We notice that at the doping concentration of x=0 and x=0.5,
the doped ruthenates are found to be orbital ordering. Therefore,
the two successive MIT should be accompanied by the transitions of
the ordered orbital states, which have not been well treated in the
literature. To account for the properties of these two MIT, the
orbital ordering should be taken into account properly within the
DMFT approach, which has gone beyond the scope of this paper.

\section{Conclusions}
\label{con}

We have studied the effects of two important factors, i.e. the
crystal field splitting and interorbital hopping, on OSMT in
nondegenerate two-orbital models and in the compounds
Ca$_{2-x}$Sr$_{x}$RuO$_{4}$. Our results show that the negative
crystal field splitting can hybridize the two nondegenerate orbitals
at the same lattice site, and strongly restrain the appearance of
the OSMT phase. However, the finite interorbital hopping
considerably expands the OSMT region in the phase diagram. In the
doping region form $x=0.2$ to 2.0 in Ca$_{2-x}$Sr$_{x}$RuO$_{4}$,
the crystal field splitting introduced by the distortion of the
RuO$_{6}$ octahedra is found to be negative. Therefore, the
occurrence of OSMT in compound Ca$_{2-x}$Sr$_{x}$RuO$_{4}$ unlikely
happens in the corresponding doping region.

\section*{Acknowledgments}
Authors thank useful discussion with A. Liebsch and Z. Fang. The
work was supported by the project-sponsored by SRF for ROCS, SEM,
the NSFC of China, no. 10874186, National Basic Research Program of
China (Grant No. 2007CB925004), the BaiRen Project, and the
Knowledge Innovation Program of Chinese Academy of Sciences.


\begin{thebibliography}{}
\bibitem{Nagaosa}
Y. Tokura and N. Nagaosa, Science {\bf 288}, 462 (2000).

\bibitem{Anisimov}
V. I. Anisimov, I. A. Nekrasov, D. E. Kondakov, T. M. Rice, and M.
Sigrist,
Eur. Phys. J. B {\bf 25}, 191 (2002).

\bibitem{Nakatsuji}
S. Nakatsuji and Y. Maeno, Phys. Rev. Lett. {\bf 84}, 2666 (2000);
Phys. Rev. B {\bf 62}, 6458 (2000).

\bibitem{Laad}
M. S. Laad, L. Craco and E. Muller-Hartmann,
Phys. Rev. B {\bf 73}, 045109 (2006); M. S. Laad, L. Craco and
E.Muller-Hartmann, Europhys. Lett. {\bf 69}, 984 (2005).

\bibitem{Lee}
J. S. Lee, S. J. Moon, T. W. Noh, S. Nakatsuji, and Y. Maeno,
Phys. Rev. Lett. {\bf 96}, 057401 (2006).

\bibitem{Wang}
S.-C. Wang, H.-B. Yang, A. K. P. Sekharan, S. Souma, H. Matsui, T.
Sato, T. Takahashi, C.-X. Lu, J.-D. Zhang, R. Jin, D. Mandrus, E. W.
Plummer, Z. Wang, and H. Ding,
Phys. Rev. Lett. {\bf 93}, 177007 (2004).

\bibitem{Neupane}
M. Neupane, P. Richard, Z.-H. Pan, Y. Xu, R. Jin, D. Mandrus, X.
Dai, Z. Fang, Z. Wang, and H. Ding, arXiv:0808.0346 (2008).

\bibitem{Koga}
A. Koga, N. Kawakami, T. M. Rice, and M. Sigrist, Phys. Rev. Lett.
{\bf 92}, 216402 (2004); Phys. Rev. B {\bf 72}, 045128 (2005); A.
Koga, K. Inaba, and N. Kawakami, Prog. Theo. Phys. Suppl. {\bf 160},
253 (2005); K. Inaba and A. Koga, Phys. Rev. B {\bf 73}, 155106
(2006).

\bibitem{Liebsch}
A. Liebsch and T. A. Costi, Eur. Phys. J. B {\bf 51}, 523 (2006); A.
Liebsch, Phys. Rev. Lett. {\bf 95}, 116402 (2005).


\bibitem{Knecht}
C. Knecht, N. Bl\"{u}mer, and P. G. J. van Dongen,
Phys. Rev. B {\bf 72}, R081103 (2005).

\bibitem{Medici}
L. de' Medici, A. Georges, and S. Biermann,
Phys. Rev. B {\bf 72}, 205124 (2005).


\bibitem{Arita}
R. Arita and K. Held, Phys. Rev. B {\bf 72}, 201102(R) (2005).

\bibitem{Ferrero}
M. Ferrero, F. Becca, M. Fabrizio, and M. Capone,
Phys. Rev. B {\bf 72}, 205126 (2005).

\bibitem{Ruegg}
A. R\"{u}egg, M. Indergand, S. Pilgram, M. Sigrist,
Eur. Phys. J. B {\bf 48}, 55 (2005).


\bibitem{Dai}
X. Dai, G. Kotliar, and Z. Fang,
arXiv: cond-mat/0611075 (2006).

\bibitem{Liebsch2}
A. Liebsch and H. Ishida, Phys. Rev. Lett. {\bf 98}, 216403 (2007).

\bibitem{Buenemann}
J. Buenemann, D. Rasch, and F. Gebhardet, J. Phys. Cond. Matt. {\bf
19}, 436206 (2007).

\bibitem{Fang0}
Z. Fang and K. Terakura, Phys. Rev. B {\bf 64}, R020509 (2001).


\bibitem{Potthoff}
M. Potthoff, Phys. Rev. B {\bf 64}, 165114 (2001).

\bibitem{Song} Y. Song and L.-J. Zou, Phys. Rev. B {\bf 72}, 085114 (2005).

\bibitem{Georges}
A. Georges, G. Kotliar, W. Krauth, and J. Rozenberg,
Rev. Mod. Phys. {\bf 68}, 13 (1996).

\bibitem{Bulla}
R. Bulla, T. A. Costi, and T. Pruschke, Rev. Mod. Phys. {\bf 80},
395 (2008).

\bibitem{Rozenberg}
M. J. Rozenberg, G. Kotliar, H. Kajueter, G. A. Thomas, D. H.
Rapkine, J. M. Honig, and P. Metcalf,
Phys. Rev. Lett. {\bf 75}, 105 (1995).

\bibitem{Sugano}
S. Sugano, {\it et al.}, "Multiplets of Transition-Metal Ions in
Crystals",  Chapt.1, Academic Press, New York (1970).

\bibitem{Song2}
Y. Song and L.-J. Zou, (unpublished).

\bibitem{Freidt}
O. Friedt, M. Braden, G. Andre, P. Adelmann, S. Nakatsuji, Y. Maeno,
Phys. Rev. B {\bf 63}, 174432 (2001).

\bibitem{Millis}
P. Werner and A. J. Millis, Phys. Rev. Lett. {\bf 99}, 126405
(2007).

\end{thebibliography}
\end{document}